**Development of quantum perspectives in modern physics**

Charles Baily[*] and Noah D. Finkelstein
Department of Physics, University of Colorado, Boulder CO 80309-0390

**ABSTRACT:** Introductory undergraduate courses in classical physics stress a perspective that can be characterized as *realist*; from this perspective, all physical properties of a classical system can be simultaneously specified and thus determined at all future times. Such a perspective can be problematic for introductory quantum physics students, who must develop new perspectives in order to properly interpret what it means to have knowledge of quantum systems. We document this evolution in student thinking in part through pre- and post-instruction evaluations using the Colorado Learning Attitudes about Science Survey. We further characterize variations in student epistemic and ontological commitments by examining responses to two essay questions, coupled with responses to supplemental quantum attitude statements. We find that, after instruction in modern physics, many students are still exhibiting a realist perspective in contexts where a quantum-mechanical perspective is needed. We further find that this effect can be significantly influenced by instruction, where we observe variations for courses with differing learning goals. We also note that students generally do not employ either a realist or a quantum perspective in a consistent manner.

**CONTENTS:**

[*] Email: Charles.Baily@Colorado.EDU

## I. INTRODUCTION

In the last decade, studies of student beliefs about physics have become a focus of interest in the physics education research (PER) community. Several assessment instruments have been developed in order to characterize student beliefs about the nature of physics and of learning physics, [1-4] including the Colorado Learning Attitudes about Science Survey (CLASS). [5] Previous studies of introductory physics courses have used the CLASS to show that student beliefs can be correlated with conceptual understanding, [6, 7] as well as with self-reported interest in physics. [8] With notably few exceptions, [1, 9, 10] studies have found that students tend to shift to more unfavorable (novicelike) beliefs about physics and of learning physics. [1, 7] However, relatively little attention has been paid to student beliefs about physics beyond introductory courses in classical mechanics and electrodynamics. [11] We seek to examine the development of student beliefs as they make key transitions from learning introductory classical physics to their first advanced or specialized course in modern physics.

Prior research on modern physics [12-14] has been predominately concerned with identifying student misconceptions and difficulties in learning the formalism of quantum mechanics. Surveys have been developed to assess students of quantum physics, but have generally focused on common difficulties for advanced undergraduate and beginning graduate students, such as the calculation of expectation values or the time evolution of a quantum state, [15, 16] or they have studied how students interpret physical meaning from graphical representations of various wave functions. [17] Others have developed conceptual surveys appropriate for lower-division modern physics students based on research on common student misconceptions. [18-21] Still, student commitments with respect to ontology (mental models of the physical world) and epistemology (beliefs about the nature of knowing) in the context of quantum physics have been understudied, particularly regarding the potentially difficult transition students make from learning classical physics to learning quantum physics.

Introductory courses in classical physics promote a perspective that we call

*local realism*. A realist perspective is deterministic, in the sense that all physical quantities describing a system can be simultaneously specified and accurately predicted for all future times. Such a perspective is often employed in the context of classical electrodynamics; for example, students are typically instructed to model an electron as a localized particle having both a well-defined position and momentum. This idea of *locality* can sometimes be useful in the context of modern physics: when learning about the photoelectric effect, a particle model for both electrons and photons is required. A particle model would be inappropriate, however, when trying to explain the interference pattern seen in a double-slit diffraction experiment. In this case, from a *quantum-mechanical perspective*, electrons and photons behave as delocalized waves as they propagate through space and as particles when interacting with a detector. A quantum perspective also recognizes the probabilistic nature of measurements performed on quantum-mechanical systems, in contrast to the determinism assumed by Newtonian mechanics. [22]

This paper concerns itself with how student perspectives change as they make the transition from learning classical physics to learning quantum physics. An analysis of student responses to pre- and post-instruction surveys at various stages of an undergraduate introductory sequence allows us to infer the development and reinforcement of a deterministic perspective during classical physics instruction, as well as the emergence of a probabilistic perspective in students as they progress through a course in modern physics. In addition, from comparative studies of two classes, we find that student perspectives can be significantly influenced by an instructor's choice of learning goals. Students are more likely to apply a quantum perspective following a course where such a perspective is explicitly taught. We also demonstrate that a student's degree of commitment to either a realist or a quantum perspective is not necessarily robust across contexts. We find that students may simultaneously hold both realist and quantum perspectives and not always know when to employ each of these epistemological and ontological frames. We conclude from the available data that specific attention paid to the ontological interpretation of quantum processes during instruction may aid students in the cultivation of a desired quantum perspective.

## II. STUDIES

The University of Colorado offers a three-semester sequence of calculus-based introductory physics: PHYS1 and PHYS2 are large-lecture courses ($N \sim 300$-$600$) in classical mechanics and electrodynamics, respectively, [23] and PHYS3 is a course in modern physics, offered in two sections, each with a typical class size of $\sim 75$ students.  At the beginning and end of each semester, students from each of the above courses were asked to respond to a series of survey questions designed to probe their epistemic and ontological commitments.  The first of these surveys was an online version of the CLASS, wherein students responded using a five-point Likert-type scale (ranging from strong disagreement to strong agreement) to a series of 42 statements, including:

> **Number 41:** "It is possible for physicists to carefully perform the same experiment and get two very different results that are both correct."

Responses to this statement are not scored by CLASS researchers because there is no consensus among experts as to how to respond. [5] The statement's ambiguities allow for a number of legitimate interpretations to emerge when formulating a response: expert physicists may disagree on what it means to conduct the "same" experiment, what qualify as "very different" results, or even what it means for an experimental result to be considered "correct."  In informal interviews, faculty members at the University of Colorado responded approximately 35% agree, 5% neutral, and 60% disagree.

### II.A. Student ideas about measurement change over time

There is a clear trend in how student responses to statement No. 41 change over the course of this introductory sequence.  In a cross-sectional study of three introductory courses in physics (PHYS1, $N$ =2200; PHYS2, $N$ =1650; and PHYS3, $N$ =730) we see a shift first from agreement to disagreement and then back to agreement with this statement. [24] Among students starting off in PHYS1, many more will agree (40%) with this statement than disagree (26%); yet the number in agreement decreases significantly following instruction in classical physics (to 30%,

*p* < 0.001), while an increasing number of students disagree (39%). This trend then reverses itself over a single semester of modern physics; at the end of which a greater percentage of students agree with this statement (46%) than at the beginning of classical physics instruction. In a longitudinal study of 124 students over three semesters, we observe the same trends, shown in Fig. **1**. Student responses at the end of the sequence are statistically indistinguishable from those at the beginning (in terms of agreement or disagreement), and so we investigate the reasons offered by students in defense of their responses in order to identify if their reasoning had changed.

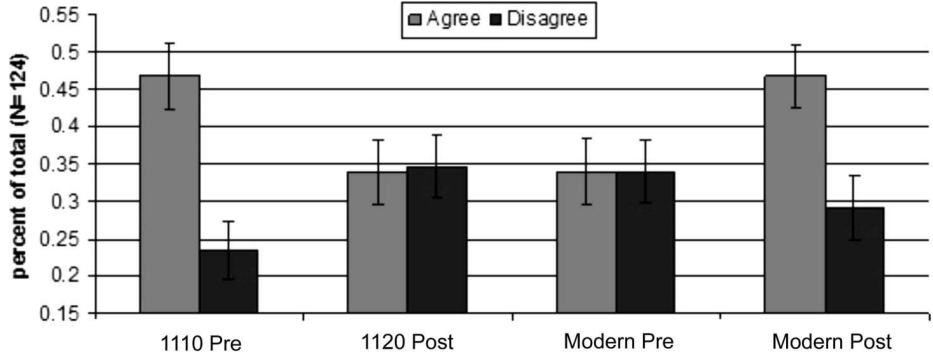

**FIG. 1.** A longitudinal study of responses to CLASS No. 41 for 124 students across a sequence of courses in introductory physics. "1110" = PHYS1; "1120" = PHYS2; "Modern" = PHYS3.

In order to clarify and gain insight into students' interpretations of statement No. 41, we analyzed the reasoning provided by approximately 600 students in an optional text box following the multiple choice response. These open-ended responses were coded into five categories through an emergent coding scheme. [25] [Table **I**] The distribution of responses for pre-instruction modern physics students was similar to that of classical physics students, and so the data for both have been combined into a single group, shown in Table **II**.

**TABLE I.** Categorization of concepts invoked by students in response to CLASS No. 41.

| A | Quantum theory or phenomena |
|---|---|
| B | Relativity or different frames of reference |
| C | There can be more than one correct answer to a physics problem<br>Experimental results are open to interpretation |
| D | Experimental, random, or human error<br>Hidden variables or chaotic systems |
| E | There can be only one correct answer to a physics problem<br>Experimental results should be repeatable |

Our analysis shows that, prior to instruction in modern physics, 59% of those who agreed with statement No. 41 offered category D explanations (experimental error or hidden variables), while category E (physics problems have only one correct answer) was preferred by those who disagreed (69%). We conclude from this that, among students of introductory classical physics, those who disagree with No. 41 primarily concern themselves with the idea that there can be only one correct result for any physical measurement, while those who agree with the statement are more conscious of the possibility for random hidden variables to influence the outcomes of two otherwise identical experiments. We find that few students invoke quantum phenomena when responding before any formal instruction in modern physics (despite the fact that a majority of these modern physics students reported having heard about quantum mechanics in popular venues, such as books by Greene [26] and Hawking, [27] before enrolling in the course); however, a single semester of modern physics instruction results in a significant increase in the percentage of students who believe that quantum phenomena could allow for two valid, but different, experimental results. Students shift from 10% to 32% in providing quantum-specific reasoning, and from 13% to 49% in referencing quantum or relativistic reasons for agreeing with the statement. [Table **III**] Responses from each population were compared with a chi-square test and were found to be statistically different ($p < 0.001$).

| TABLE II. Distribution of reasons invoked by students in response to CLASS No. 41 **before** instruction in modern physics. | | |
|---|---|---|
| $N = 507$ | DISAGREE (%, $N=199$) | AGREE (%, $N=231$) |
| A | 5 | 10 |
| B | 0 | 3 |
| C | 6 | 28 |
| D | 20 | 59 |
| E | 69 | 0 |
| TOTAL | 100 | 100 |

| TABLE III. Distribution of reasons invoked by students in response to CLASS No. 41 **after** instruction in modern physics. | | |
|---|---|---|
| $N = 83$ | DISAGREE (%, $N=26$) | AGREE (%, $N=41$) |
| A | 27 | 32 |
| B | 4 | 17 |
| C | 8 | 10 |
| E | 19 | 41 |
| D | 42 | 0 |
| TOTAL | 100 | 100 |

**II.B. Influence of instruction on student perspectives**

Students' commitments to either a realist or quantum perspective can vary by context; [28] to see if these commitments can be influenced by different types of instruction and learning goals, we examined data from two recent semesters of PHYS3 for physics majors. Course PHYS3A was taught by a PER instructor who employed in-class research-based reforms, [29] including computer simulations [30] designed to provide students with a visualization of quantum processes. Course PHYS3B was taught the following semester in the form of traditional lectures delivered from a chalkboard. Both classes provided online and written homework, standard 1-h exams, and a final exam. A typical semester of modern physics at the University of Colorado devotes roughly one third of the course to special relativity, with the remaining lectures covering the foundations of quantum mechanics and simple applications. A notable difference in these two courses is the instructional perspectives and learning goals of the two instructors. Through informal end-of-term interviews, and an analysis of posted solutions to survey questions related to measurement and uncertainty, it is clear that the instructors held different beliefs about the role of introducing quantum measurement when teaching modern physics. In the context of a double-slit diffraction experiment, the instructor for PHYS3A explicitly taught students to think of the electron as a delocalized wave that passes through both slits and interferes with itself, and then becomes localized upon measurement; the instructor for PHYS3B preferred a more agnostic stance on how to think of the electron between being emitted and being detected, and generally did not address such issues.  Despite instructor B's agnostic perspective, when posting solutions to the quantum mechanics conceptual survey, [19] he instructed students that each electron went through either one slit or the other, but that it is fundamentally impossible to determine which one without destroying the interference pattern.

Students from both of these courses were given two end-of-term essay questions.  The first of which asked them to argue for or against statements made by three fictional students who discuss the representation of an electron in the quantum wave interference (QWI) PhET simulation. [31; see Fig. **2**] In this simulation, a large bright circular spot representing the probability density for a single electron (A) emerges from a gun, (B) passes through two slits, and (C) a small dot appears on a detection screen; after a long time (many electrons) an interference pattern develops (not shown).  Each of the following statements made by a fictional student is meant to represent a potential perspective on how to model the electron between the time it is emitted from the electron gun and when it is detected at the screen.

**Student 1:** "That blob represents the probability density, so it tells you the probability of where the electron could have been before it hit the screen.  We don't know where it was in that blob, but it must have actually been a tiny particle that was traveling in the direction it ended up, somewhere within that blob."

**Student 2:** "No, the electron isn't inside the blob, the blob represents the electron! It's not just that we don't know where it is, but that it isn't in any one place.  It's really spread out over that large area up until it hits the screen."

**Student 3:** "Quantum mechanics says we'll never know for certain, so you can't ever say anything at all about where the electron is before it hits the screen."

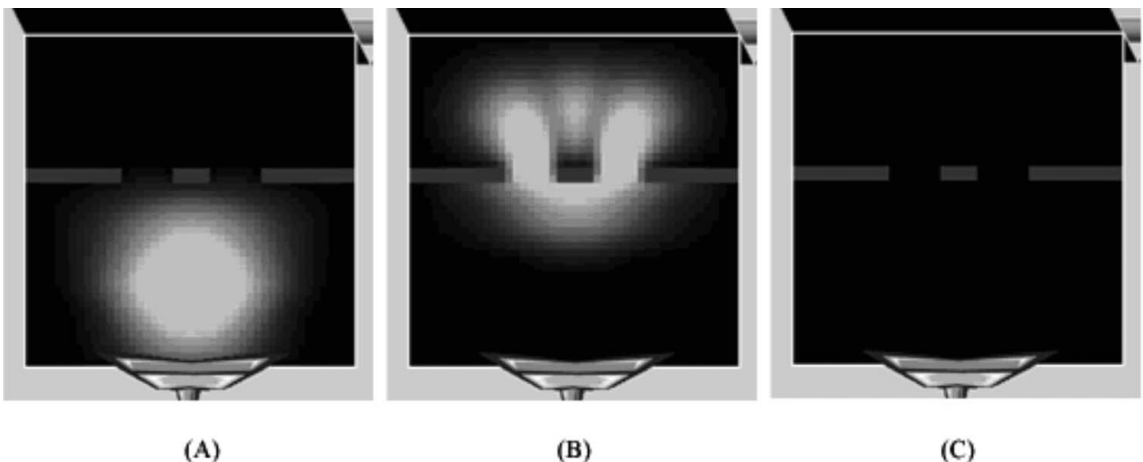

FIG. 2. A sequence of screenshots from the quantum wave interference PhET simulation.

In this end-of-term survey question, students were asked to agree or disagree with any or all of the fictional students and to provide evidence in support of their response. Responses were coded according to whether students preferred a realist or a quantum perspective in their argumentation. A random sample of 20 student responses were recoded by a PER researcher unaffiliated with this project as a test for inter-rater reliability; following the discussion of the coding scheme, the two codings were in 100% agreement. The following quotations from two students are illustrative of the types of responses seen:

**Student response (realist):** "We just can't know EXACTLY where the electron is and thus the blob actually represents the probability density of that electron. In the end, only a single dot appears on the screen; thus the electron, wherever it was in the probability density cloud, traveled in its own direction to where it ended up."

**Student response (quantum):** "The blob is the electron and an electron is a wave packet that will spread out over time. The electron acts as a wave and will go through both slits and interfere with itself. This is why a distinct interference pattern will show up on the screen after shooting out electrons for a period of time."

The distribution of all responses for the two courses is summarized in Table **IV**; columns do not add to 100% because some students provided a mixed or otherwise unclassifiable response. For this essay question, there is a strong bias toward a quantum perspective among PHYS3A students, while students from PHYS3B highly preferred a realist perspective. Notably, virtually no student agreed with fictional student 3 (which would be consistent with an agnostic perspective); among those who explicitly disagreed with student 3, most felt that knowing the probability density was a sufficient form of knowledge about this quantum system.

**TABLE IV.** Student response to the quantum wave interference essay question from two recent semesters of PHYS3. Numbers in parentheses represent the standard error (in percent) on the proportion.

|  | PHYS3A ($N = 72$) | PHYS3B ($N = 44$) |
|---|---|---|
| Realist | 18 (5) | 75 (7) |
| Quantum | 78 (5) | 11 (5) |

A second essay question sought to examine students' ideas about the notion of uncertainty in classical systems and their relation to quantum phenomena. This question concerns a *Plinko* game, consisting of a marble and a board with a number of pegs (see Fig. **3**). When the marble is released from its starting position it encounters a series of pegs as it falls and is deflected either to the left or to the right each time it hits a peg, until it ends up in one of the 12 slots at the bottom of the board. The text of the question reads as follows:

> Suppose a machine releases a marble from the same starting point 300 times, and the cumulative results for where the marble ends up are shown in the histogram below (Fig. **3**). There is a distribution of possible final outcomes for each drop of the marble, even though the initial conditions for each drop seemed to be the same.
>
> Q2-(I) What is the origin of the uncertainty in the final outcome for this classical system? (Explain your answer in 2–3 sentences at most.)
>
> Q2-(II) The distribution shown in the histogram above looks similar to a distribution of measurements on a quantum system (for example, one part of an interference pattern created during a double-slit experiment). In what ways is the uncertainty in final outcomes for such a quantum system the same as or different from the classical example given above?

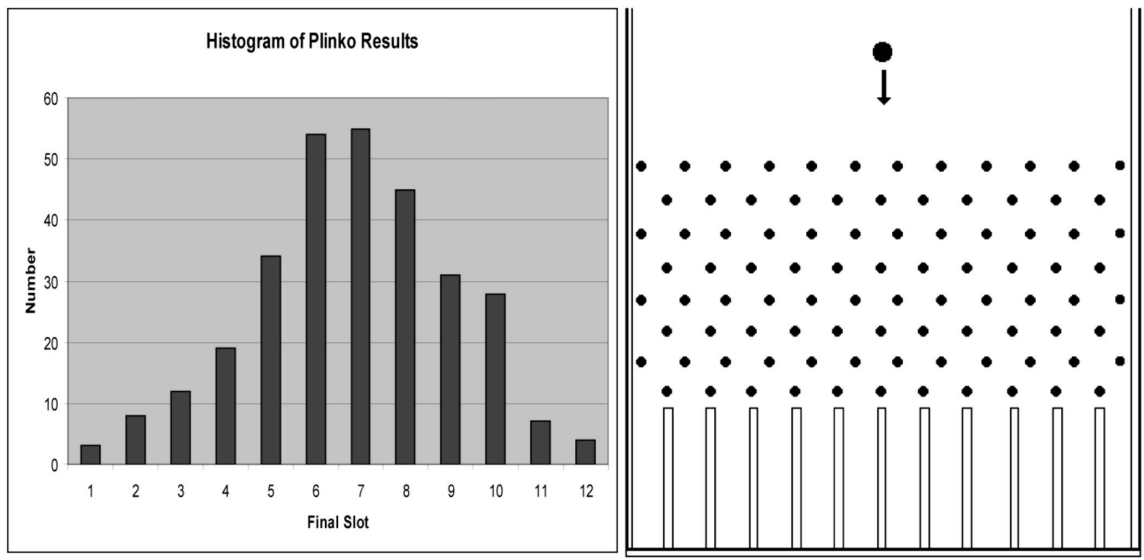

**FIG. 3.** A depiction of a Plinko game and a histogram of results from a series of 300 releases of the marble from the same starting position.

This question was designed to cue students to consider how an uncertainty in initial conditions will lead to varying final outcomes in a classical system (specifically, through the phrase "the initial conditions for each drop seemed to be the same"). Students were then asked to compare and contrast this origin of uncertainty (for a chaotic but deterministic system) with an example from quantum physics. We expect that students who are committed to a realist perspective would view the two examples as similar, so that the uncertainty in initial conditions for the *Plinko* marble would be seen as analogous to the perceived uncertainty in the initial conditions for electrons in a diffraction experiment. Most every student provided a satisfactory response to the first part of the question, stating either that the distribution of final results was due to a random 50/50 probability on how the marble would be deflected at each peg or due to an uncertainty in the initial conditions for this classical system. Student responses to part II were rated using the rubric shown in Table **V**; most every student commented on the systems being similar in that there is some kind of distribution of final results (or that both can be described with statistics, probability, etc.), and so the rubric does not code for this response and focuses instead on the argued differences between the two systems. The results for all student responses are summarized in Table **VI**.

TABLE V. Categorization of responses to part II of the Plinko essay question.

| | |
|---|---|
| A (Quantum) | Different because uncertainty is inherent to quantum systems. Measurements on identical QM systems can have varying outcomes. |
| B (Quantum) | Different because there is no interference or the marble is localized in space. "Electrons behave like waves." |
| C (Realist) | No statement about differences or thinks they are the same. Implies there are differences, but reasoning is unclear or weak |

TABLE VI. Results for Q2-(II) for two recent semesters of PHYS3.

| | | | | | |
|---|---|---|---|---|---|
| PHYS3A ($N=70$) | A | 13% | PHYS3B ($N=44$) | A | 18% |
| | B | 49% | | B | 20% |
| | C | 38% | | C | 61% |

The following is the full response of one student who coded as an A category:

Q2-(I) "The origin for the uncertainty comes from the variables of the initial conditions. The Plinko ball can't be dropped exactly the same way every time, and so not all the balls follow the same path."

Q2-(II) "In a quantum system, the initial conditions can be exactly the same in every case, but the outcomes can be different. The reason the quantum distribution looks the same as the macro distribution is because quantum distributions follow probabilities which are similar to classical distribution patterns."

As can be seen in Table **VI**, few students from either semester provided the complete targeted response, which was to recognize that the classical Plinko game is a false analogy to a quantum system, where there can be varying outcomes to measurements even though the initial conditions are identical. Still, a majority of PHYS3A students perceived that there is some difference between the two examples (the most common response is that the classical system does not exhibit interference effects or that electrons behave as waves, while the marble does not), while a majority of PHYS3B students seemed to believe that the origins of uncertainty in both systems were analogous or were unable to articulate why they might be different.

Students from both PHYS3 courses also responded at the beginning and end of the semester to additional statements appended to an online version of the CLASS for modern physics students, including:

**QA No. 16:** "An electron in an atom has a definite but unknown position at each moment in time."

It might be expected that a student who has learned to view an electron as being delocalized in space in the context of an electron diffraction experiment should also see it as such when considering whether an electron in an atom can have a definite position in the absence of measurement. While we again observe differences in the two course offerings, Table **VII** shows there is no strong bias toward a single perspective as was seen in Table **IV**. From a quantum perspective, disagreement with QA No. 16 can be characterized as favorable. Table **VII** shows that students in PHYS3A posted a 22-point increase in favorable responses and those from PHYS3B posted a 13% favorable shift; but while PHYS3A showed a 5% decrease in unfavorable responses, PHYS3B students increased their unfavorable responses at the end of the semester by 6% points.

**TABLE VII.** Student responses to the statement "An electron in an atom has a definite but unknown position at each moment in time."

|  | PHYS3A ($N = 41$) | | PHYS3B ($N = 36$) | |
|---|---|---|---|---|
|  | PRE (%) | POST (%) | PRE (%) | POST (%) |
| DISAGREE | 22 | 44 | 10 | 23 |
| NEUTRAL | 32 | 17 | 39 | 21 |
| AGREE | 44 | 39 | 48 | 54 |

## II.C. Consistency of student responses

While student responses can vary in aggregate from course to course, we investigate the robustness or consistency of student beliefs. Do students hold consistent perspectives within and across domains? We note that the degree of difference in perspectives for PHYS3A and PHYS3B is smaller for QA No. 16

(Table **VII**) than the quantum wave interference question (Table **IV**). The post-data in Table **VII** from the two courses can be combined and compared with student responses to the prior essay question on double-slit interference. Table **VIII** shows the student post-instructional responses to QA No. 16, categorized by which perspective they held on the QWI essay question of Fig. **2**. Here, we see that students who had preferred a quantum perspective tended to answer QA No. 16 favorably, while the majority of students who preferred a realist perspective chose an unfavorable response. Of particular interest, however, is that students were not necessarily consistent in their responses: 18% of those who disagreed with QA No. 16, and 33% of those who agreed, were offering a response that was inconsistent with their response to the QWI essay question. That is, 18% of students held a quantum perspective on electron position (QA No. 16), but a realist perspective on the quantum wave interference question. 33% of students were the reverse: holding a realist perspective on electron position in QA No. 16 (agree), but a quantum perspective on the interference question.

**TABLE VIII.** Combined post-responses to QA No. 16 (columns), grouped according to student responses to the quantum wave interference essay question (rows).

|         | Disagree (%) | Neutral (%) | Agree (%) | Total (%) |
|---------|--------------|-------------|-----------|-----------|
| Quantum | 56           | 11          | 33        | 100       |
| Realist | 18           | 18          | 64        | 100       |

Nonetheless, the majority of students held a consistent quantum or realist perspective on the two questions relating to electrons (quantum wave interference and electron position in an atom). By including analysis of student perspectives with respect to the second essay question (sources of classical and quantum uncertainties), we may consider the consistency of student perspective across a third context. The data for both courses (PHYS3A and PHYS3B) have been combined in Table **IX**, and student responses are categorized as quantum (or realist) if students consistently report an answer coded as quantum (or realist) across all three questions; otherwise, students are reported as mixed (giving at least one quantum and one realist answer). Table **IX** shows that the majority of students end up with a mixed perspective ($p < 0.01$ by pairwise proportion test to each of quantum and realist groups), sometimes applying a quantum perspective and sometimes a realist perspective. Intriguingly, more students ($p = 0.05$, two tailed, and pairwise) end with a consistently realist perspective than a quantum perspective. The dominance of the mixed state holds for each of the PHYS3A and PHYS3B courses independently (71% and 50%, respectively); however, in PHYS3A the quantum state is more prevalent than realist (22% vs 7%) and in PHYS3B the realist state is more prevalent than quantum (39% vs 11%).

**TABLE IX.** Consistency of student responses from both PHYS3 courses to both essay questions and QA No. 16.

| PHYS3A & B | $N = 77$ | Percentage (%) |
|---|---|---|
| All quantum | 13 | 17 |
| Mixed | 47 | 61 |
| All realist | 17 | 22 |

## III. DISCUSSION AND CONCLUSIONS

The data presented in this paper serve as evidence in support of three key findings. First, student perspectives with respect to measurement and determinism in the contexts of classical physics and quantum mechanics evolve over time. The distribution of reasoning provided by students in response to the CLASS survey statement indicates that the majority of those who disagree with this statement believe that experimental results should be repeatable or that there can be only one correct answer to a physics problem. One could easily imagine that students begin their study of classical physics at the university level with a far more deterministic view of science than is evidenced by their initial responses to the survey statement (after all, most students do arrive with some training in classical science). We take the first significant shift in student responses (a decrease in agreement and an increase in disagreement with this statement, as shown in Fig. **1**) to be indicative of the promotion and reinforcement of a deterministic perspective in students as a result of instruction in classical physics. After a course in modern physics, student responses shift a second time (an increase in agreement and a decrease in disagreement with the survey statement), although the reasoning behind their responses changed. Students of modern physics are instructed that different frames of reference could lead to different experimental results, both of which are correct. They also learn that the quantum-mechanical description of nature is probabilistic, and that the determinism assumed by Newtonian mechanics is no longer valid at the atomic scale. The influence of this type of instruction is reflected in the increase in the number of students who invoke relativistic or quantum phenomena as a reason for agreeing with the survey statement.

Second, we observe that how students develop and apply a quantum or realist perspective can depend on the instructional approach, learning goals, and tools used for teaching students. The results for the quantum wave interference essay question indicate that how students view an electron within the context of a double-slit experiment can be significantly influenced by instruction. Instructor A explicitly taught students that each electron passes through both slits and interferes with itself, and used the PhET simulation to provide students with an in-class visualization of this process. It should be noted, however, that this interpretation is not universally accepted among expert physicists. An alternative point of view insists that one cannot ask about that which cannot be known, since there is no way of determining the actual path of each electron without destroying the interference pattern. From this "agnostic" perspective, quantum mechanics concerns itself only with predicting experimental results, and the question of which slit the electron passed through is considered to be ill posed and anyway irrelevant to the application of the mathematical formalism. Although Instructor B reported holding such an agnostic stance, he did not teach this perspective explicitly, and virtually none of his students applied an agnostic perspective when responding to the quantum wave interference essay question; instead, a majority of PHYS3B students applied a realist interpretation.

While we do not make a valuation of either of these instructional goals, we feel it is worth acknowledging that different goals regarding the interpretation of quantum processes do exist. We believe that the physics community would benefit from a discussion of the pedagogical usefulness of either of these interpretations because our research indicates that students, in this regard, can adopt their instructor's philosophical predisposition when given explicit instruction. We believe that this in itself is a significant finding, considering that there is substantial evidence that students do not necessarily adopt an instructor's views and attitudes in other contexts. For example, students will often not develop a sound conceptual understanding of physics, even if instructors believe in the importance of such, unless conceptual understanding is explicitly taught, as is evidenced by myriad studies. Furthermore, students tend not to develop more sophisticated views on the nature of science and of learning physics, even from reformed instruction in introductory courses. [1, 7] In fact, students' views on the nature of physics and learning tend to become less "expertlike" over time, although it has been shown that this trend can be positively influenced by making epistemology an explicit aspect of instruction in introductory physics courses. [10] The results of this study provide further indication that instructors should not take for granted that students will adopt their perspectives on physics unless such learning goals are made explicit in their teaching.

Third, we find that most students do not exhibit a consistent perspective on uncertainty and measurement across multiple contexts. While the data shown in Table **VIII** do demonstrate some consistency of responses when answering two questions on electron position, we see that a significant number of students who preferred the quantum description of an electron in a diffraction experiment would still agree that an electron in an atom has a definite, but unknown, position. When looking across more varied contexts that include a question comparing electron diffraction and a Plinko game, students exhibit a tendency to be less consistent, dominantly holding mixed quantum and realist perspectives. Students likely do not have a robust "concept" of quantum measurement. These findings parallel studies of student epistemic commitment in classical physics [32] and the resources view of student conceptual understanding and understanding the nature of knowing physics. [33]

In the end, we believe that a reasonable instructional objective is for students to use the appropriate perspective (deterministic or probabilistic, localized or delocalized) at the appropriate time. This goal seems to require a level of metacognitive awareness that students may not have at the introductory level: we find that few students from either course were able to demonstrate the ability to distinguish between classical uncertainty and the uncertainty that is inherent to quantum systems. While a majority of students from the transformed PHYS3A course demonstrated at least partial understanding of this distinction (by focusing on interference and the wave description of electrons), a majority of PHYS3B students did not make any reasonable distinction between the two systems (which is again consistent with a realist perspective).

These findings suggest that students do not automatically develop the robust understanding of measurement, uncertainty, or metacognitive abilities that we may desire. If we are to include these goals for our classes, it is important to understand how these messages are sent to our students and what instructional practices may promote such understandings. Such investigations are the subject of current studies.

**ACKNOWLEDGEMENTS**

The authors wish to thank the University of Colorado physics faculty members who helped facilitate this research, as well as S. B. McKagan for thoughtful discussions and suggestions, and the rest of the Physics Education Research group at Colorado for their helpful insights. This work was supported in part by NSF CAREER Grant No. 0448176 and the University of Colorado.